\documentclass[aps,prc,twocolumn,superscriptaddress,nofootinbib,showpacs,showkeys,preprintnumbers]{revtex4-1}
\usepackage{graphicx}  
\usepackage{bm}        
\usepackage{amssymb}   
\usepackage{amsmath}   
\usepackage{hyperref}
\usepackage{lineno}

\usepackage{color}
\definecolor{dgreen}{cmyk}{1.,0.,1.,0.2}        
\definecolor{orange}{cmyk}{0.,0.353,1.,0.}    

\def\snn{\mbox{$\sqrt{s_{_{\rm NN}}}$}}
\def\trento{T$_{\rm R}$ENTo}
\def\AFTER{\mbox{AFTER@LHC}}

\begin{document}
\title{A benchmark of initial state models for heavy-ion collisions\\ at $\snn=$~27 and 62~GeV}
\author{Jakub Cimerman} \affiliation{Faculty of Nuclear Sciences and Physical Engineering, Czech Technical University in Prague,\\  B\v rehov\'a 7, 11519 Prague 1, Czech Republic}
\author{Iurii Karpenko} \affiliation{Faculty of Nuclear Sciences and Physical Engineering, Czech Technical University in Prague,\\  B\v rehov\'a 7, 11519 Prague 1, Czech Republic}
\author{Boris Tom\'a\v{s}ik} \affiliation{Faculty of Nuclear Sciences and Physical Engineering, Czech Technical University in Prague,\\  B\v rehov\'a 7, 11519 Prague 1, Czech Republic}
\affiliation{Univerzita Mateja Bela, Tajovsk\'eho 40, 974~01 Banská Bystrica, Slovakia}
\author{Barbara Antonina Trzeciak} \affiliation{Faculty of Nuclear Sciences and Physical Engineering, Czech Technical University in Prague,\\  B\v rehov\'a 7, 11519 Prague 1, Czech Republic}

\begin{abstract}
Description of relativistic heavy-ion collisions at the energies of RHIC Beam Energy Scan program with fluid dynamic approach poses several challenges, one of which being a complex geometry and a longer duration of the pre-hydrodynamic stage. Therefore, existing fluid dynamic models for heavy-ion collisions at the RHIC Beam Energy Scan energies rely on rather complex initial states, such as UrQMD cascade or multi-fluid dynamics. In this study, we show that functionally simpler, non-dynamical initial states can be employed for the fluid dynamical simulations of Au-Au collisions at $\snn=27$ and 62.4~GeV. We adapt the initial states based on Monte Carlo Glauber model (GLISSANDO 2) and $\sqrt{T_A T_B}$ ansatz based on reduced thickness (\trento\ $p=0$), extended into the longitudinal direction and finite baryon density.
We find that both initial states, when coupled to a 3D event-by-event viscous fluid dynamic + cascade model, result in an overall fair reproduction of basic experimental data: pseudorapidity distributions, transverse momentum spectra and elliptic flow, at both collision energies. This is a rather surprising, given that the $\sqrt{T_A T_B}$ ansatz is functionally similar to the EKRT and IP-Glasma models, which are successful at much larger energies and rely on a partonic picture of the initial state.
\end{abstract}
\pacs{25.75.Ld, 25.75.Gz, 05.70.Fh}
\maketitle

\section{Introduction}
Strongly interacting matter becomes deconfined when it is brought into a state with very high energy density. The transition is a crossover when the net baryon density vanishes and likely becomes a first-order at so-far unknown value of the baryon chemical potential. 
The first unambiguous experimental signatures of this new state of matter, the Quark-Gluon Plasma (QGP), came from the gold-gold collisions at the Relativistic Heavy Ion Collider (RHIC) at BNL. An associated signature of the strongly-interacting system with a small mean free path came in a form of observed strong elliptic flow.

Last decade brought substantial improvements to the hydrodynamic modelling of heavy-ion collisions. In particular, the variety of initial state models has expanded from geometrical Glauber picture based on nucleonic constituents, to a family of Color Glass Condensate models, most notably Kharzeev-Levin-Nardi (KLN) model \cite{Kharzeev:2001yq}, to the IP-Glasma approach \cite{Schenke:2012wb}, which combines the impact parameter dependent saturation scale and Yang-Mills evolution of the initial quasi-classical gluon fields up to the supposed start of the hydrodynamic description. It is now understood that fluctuating initial state and event-by-event fluid dynamic modelling are necessary ingredients to describe the odd Fourier harmonics of the momentum distribution of produced hadrons as a function of the azimuthal angle in the transverse plane.

The initial state of a heavy ion collision is not directly accessible to be probed with hadronic observables. Therefore, there are two strategies to constrain the properties of the initial state: either (i) to extend the list of hadronic observables which have to be described within a hydrodynamic model with a given initial state, or (ii) to find observables which would be particularly sensitive to the properties of the initial state, and less sensitive to the properties of the subsequent hydrodynamic evolution, such as the shear viscosity or the equation of state. Both strategies have been quite fruitful so far. For example, the calibration of a viscous hydrodynamic + cascade model to experimentally measured transverse momentum spectra of pions, kaons and protons, yields and flow harmonics $v_2,\dots v_4$ of all charged hadrons at $\snn=2.76$~TeV (LHC) energy using Bayesian analysis~\cite{Bernhard:2016tnd,Bernhard:2019bmu} provided a robust constraints on the initial entropy density as function of the nuclear thickness, as well as on the shear viscosity of the hydrodynamic medium. Another example is a study in the EKRT+viscous hydrodynamic framework~\cite{Niemi:2015qia}, where the temperature-dependent ratio of the shear viscosity to entropy density of the QGP medium was constrained from a simultaneous reproduction of centrality-dependent multiplicities, transverse momentum spectra, elliptic flow in AA collisions at $\snn=200$~GeV (RHIC) and $\snn=2.76$~TeV (LHC), plus various extra 2- and 3-particle correlation observables measured at $\snn=2.76$~TeV energy. An example of the strategy (ii) is a study which demonstrates that the ratios of flow harmonics measured with different cumulant methods---$v_2\{4\}/v_2\{2\}, v_3\{4\}/v_3\{2\}$---provide constraints on the structure of the initial state irrespective of the shear viscosity of the subsequent hydrodynamic evolution~\cite{Giacalone:2017uqx}. 

Most of the hydrodynamic studies for the LHC energies, including the studies above, are based on the approximation of longitudinal boost invariance. This approximation works quite well at those energies as long as mid-rapidity observables are concerned. It allows one to reduce the fluid dynamical modelling to transverse directions only, assuming scaling flow in the longitudinal (beam) direction.

Recently, in the light of the RHIC Beam Energy Scan (BES) program and related lower-energy experiments under construction at GSI FAIR and JINR NICA, there is a great interest to revise the hydrodynamic picture for collision energies from a few to a hundred GeV per nucleon-nucleon pair in the center-of-mass frame. There are several challenges in this direction, such as finite baryon and electric charge densities, absence of the longitudinal boost invariance, as well as more complex geometry of the initial state. The latter is a direct consequence of the weaker Lorentz contraction of the incoming nuclei due to the lower collision energy. Despite these challenges, a pioneering study~\cite{Karpenko:2015xea} has shown that a 3+1 dimensional event-by-event viscous hydrodynamic picture provides a generally good description of the data at lower collision energies.

However, the aforementioned study dealt with only one initial state model, which is the UrQMD cascade~\cite{Bass:1998ca}. The UrQMD cascade, for most part, is based on the hadronic degrees of freedom. At the end of the initial stage, the first row of the energy-momentum tensor $T^{0\mu}$ is used to calculate the initial energy density and flow velocities for the hydrodynamic expansion. Such forced hydrodynamization results in a non-zero pre-thermal flow, generated by interactions in the UrQMD cascade. Generally, both initial transverse flow is non-zero and the initial longitudinal flow is different from a boost-invariant scaling profile $v_z=z/t$.

Another example of a fluid dynamic model for lower energies, down to a few GeV per nucleon pair, is the 3-fluid dynamics (3FD) model~\cite{Ivanov:2005yw}, where the initial stage of heavy-ion collisions is treated as an inter-penetration of counter-flowing baryon-rich fluids. In the 3FD, the dynamics starts as soon as the fluids touch each other, or when the first nucleon-nucleon collisions take place.

It is therefore an interesting question whether one can resort to a simpler, purely geometrical initial state model such as Monte Carlo Glauber, which has been widely used for top RHIC and LHC energies, and still be able to reproduce the basic experimental observables at the BES energies in a viscous hydrodynamic model. Because of the longer duration of the pre-hydrodynamic stage at lower energies, it is also important to understand whether the pre-hydrodynamic interactions in the initial stage UrQMD are mandatory, which would make microscopic models such as UrQMD or the newly developed SMASH \cite{Weil:2016zrk} unique for the lower energy applications within the hydrodynamic approach.

In this work, we focus on two particular collision energies in the BES range, namely $\snn=27$ and 62.4~GeV, and examine the reproduction of the basic experimental data: transverse momentum spectra of the most abundant hadrons and their elliptic flow coefficients, in 3 dimensional event-by-event viscous hydrodynamic model with three different initial state models: UrQMD \cite{Bass:1998ca}, Monte Carlo Glauber (implemented via GLISSANDO 2 code~\cite{Rybczynski:2013yba}) and reduced thickness (via \trento\ code)~\cite{Moreland:2014oya}. The initial conditions from the latter two initial state models are extended into the longitudinal space-time rapidity. We describe the setup and the different initial state models in Section~\ref{sect-model}, discuss the results and the implications for the early-time dynamics at such collision energy in Section~\ref{sect-results}, and then conclude in the last Section.

\section{The model}\label{sect-model}
Modelling of the heavy-ion collision dynamics is performed in a 3-stage fashion. For the initial state, three different models are used: UrQMD, GLISSANDO 2 and \trento. In all cases, the transition from the initial state to the fluid dynamic description, or hydronynamization, takes place at $\tau=\tau_0$. The subsequent evolution of the hot and dense matter is performed with the 3 dimensional viscous code vHLLE~\cite{Karpenko:2013wva}. The transition from hydrodynamic fields to particles, or hadrons, (particlization) takes place at the hypersurface of fixed energy density $\epsilon_\text{sw}=0.5$~GeV/fm$^3$. At the particlization, the hadrons are sampled using phase-space distributions via the Cooper-Frye formula \cite{Cooper:1974mv}, extended with corrections due to the shear viscosity. The basics of the model are covered in~\cite{Karpenko:2015xea}.


\subsection{3D initial state scenarios}\label{subsect-IS}

For this study, we have extended the vHLLE+UrQMD model with two new 3 dimensional initial states (IS), based on the Monte Carlo Glauber approach (via GLISSANDO 2 code) and \textit{reduced thickness} ansatz (via \trento\ code) in addition to the UrQMD initial state used previously. For the reasons of brevity, further on we use the names of the codes (GLISSANDO, \trento\ and UrQMD) to refer to the initial state models. Let us describe all 3 initial state (IS) scenarios in the following.

1. At the collision energies of interest, the \textbf{UrQMD IS} invokes PYTHIA6 to simulate the initial nucleon-nucleon scatterings within the nucleus-nucleus reaction. PYTHIA6, in turn, treats the inelastic $NN$ scatterings via string formation, followed by string break-up (fragmentation). The products of the string break-up are hadrons which after their formation time are allowed to rescatter. Such secondary rescatterings generate collective expansion from the early (pre-hydrodynamic) stage of heavy-ion collisions. To perform the hydrodynamization at $\tau=\tau_0$ with UrQMD IS ($\tau = \sqrt{t^2-z^2}$ is the longitudinal proper time), the hadrons crossing the $\tau=\tau_0$ hypersurface are recorded, and hadron scatterings after $\tau=\tau_0$ are forbidden in the code. The energy and momenta of the hadrons at the $\tau=\tau_0$ hypersurface are smoothly distributed to the hydrodynamic grid using Gaussian profiles:
\begin{align} \label{urqmd-IS-Gauss}
\Delta P^\alpha_{ijk} & = P^\alpha \cdot C\cdot\exp\left(-\frac{\Delta x_i^2+\Delta y_j^2}{R_\perp^2}-\frac{\Delta\eta_k^2}{R_\eta^2}\gamma_\eta^2 \tau_0^2\right) \\
\Delta N^0_{ijk}&=N^0 \cdot C\cdot\exp\left(-\frac{\Delta x_i^2+\Delta y_j^2}{R_\perp^2}-\frac{\Delta\eta_k^2}{R_\eta^2}\gamma_\eta^2 \tau_0^2\right), \label{Gauss2}
\end{align}
where $\Delta x_i$, $\Delta y_j$, $\Delta \eta_k$ are the differences
between particle's position and the coordinates of the hydrodynamic
cell $\{i,j,k\}$, and $\gamma_\eta={\rm cosh}(y_p-\eta)$ is the
longitudinal Lorentz factor of the particle as seen in a frame moving
with the rapidity $\eta$. The normalization constant $C$ is calculated
from the condition that the discrete sum of the values of the Gaussian
in all neighboring cells equals one. The resulting $\Delta P^\alpha$
and $\Delta N^0$ are transformed into Milne coordinates and added
to the energy, momentum, baryon and electric charges of each hydro cell. This procedure
ensures that in the initial transition from transport to hydrodynamics
the energy, momentum, baryon and electric charges are conserved.
Note that with UrQMD IS, even at the lower RHIC BES energies, the local energy density of the dense hadronic system right before the hydrodynamization is often higher than the typical energy density corresponding to the transition from hadronic to partonic degrees of freedom according to lattice QCD.

2. Another IS scenario used, \textbf{GLISSANDO 2} \cite{Rybczynski:2013yba}, is a Monte Carlo implementation of the Glauber model of relativistic heavy-ion collisions. The code samples the positions of participant nucleons and binary scatterings in the transverse plane. Both participant nucleons and binary scatterings are assumed to be the \emph{sources} of energy or entropy depositions. Following the classic observation from \cite{Kharzeev:2000ph}, the entropy density is assumed to be distributed in accord with 
~\cite{Bozek:2012fw,Bozek:2015bha}:
\begin{multline}
s(x,y,\eta_s)= \kappa \sum_i f_\pm(\eta_s)\left[ (1-\alpha) + N^{\rm coll}_i \alpha \right]  \\
{} \times \exp\left( - \frac{(x-x_i)^2+(y-y_i)^2}{2\sigma^2}\right),
\end{multline}
where the sum goes through all participant nucleons $i$; $x,y$ are the coordinates in the transverse plane, $\eta_s=1/2\ln\left((t+z)/(t-z)\right)$ is the space-time rapidity (not to be confused with momentum space pseudo-rapidity $\eta$), $(x_i,y_i)$ are the positions of the participant nucleons, $N^{\rm coll}_i$ is the number of collisions the participant nucleon $i$ has suffered. The width of the Gaussian smearing is denoted $\sigma$, and in practice we take $\sigma=0.4$~fm. The normalization constant $\kappa$ will guarantee the correct total energy content of the fluid. The mixing factor $\alpha$ regulates the relative contributions from the participant nucleons and the binary scatterings. For our calculations we fix $\alpha=0.123$ for $\snn=27$~GeV and $\alpha=0.132$ for $\snn=62.4$~GeV, taken from an interpolation between the values at $\snn=19.6$ and $200$~GeV assuming logarithmic dependence of $\alpha(\snn)$. The latter values are fitted from the centrality dependence of pseudorapidity density of charged hadrons.

As GLISSANDO 2 is based on the Glauber model, it only provides the distributions in the transverse plane. Then, to construct a full 3-dimensional initial state for the hydrodynamic stage, again following~\cite{Bozek:2012fw,Bozek:2015bha}, we impose an approximate triangular shape of the space-time rapidity distribution of the entropy deposition from the forward-going (+) and backward-going (--) participant nucleons:
\begin{align}
f_{\pm}(\eta_s)&=\frac{\eta_{\rm M}\pm \eta_s}{2 \eta_{\rm M}} H(\eta_s)\ & \mbox {for } \ |\eta_s|<\eta_{\rm M}
\label{eq:lprof}
\end{align}
where the so-computed $f_\pm(\eta_s)$ is then limited to the range $[0, 1]$ so that the deposition does not become locally negative, and the finiteness of the shape in rapidity is ensured by the profile function $H(\eta_\parallel)$:
\begin{equation}
H(\eta_s)=\exp\left(-\frac{(|\eta_s|-\eta_0)^2\Theta(|\eta_s|-\eta_0)}{2\sigma_\eta^2}\right) .  \label{eq:Heta}
\end{equation}
Such rapidity profile of the entropy deposition from the participant nucleons has been successfully used in a variety of studies, such as transverse momentum correlations in Au-Au collisions at the top RHIC energy \cite{Bozek:2012fw} and Pb-Pb collision at the LHC energies \cite{Chatterjee:2017mhc}, collective flow in small systems \cite{Bozek:2011if}, or longitudinal decorrelation of flow harmonics in Pb-Pb collisions at the LHC energies \cite{Bozek:2017qir}.\par
In addition to that, we assume the baryon charge deposition from each forward- and backward-going participant into the fluid in the following form:
\begin{eqnarray}
n_B(x,y,\eta_s)&=& \kappa_B \sum_i \exp\left( -\frac{(\eta_B\pm\eta_s)^2}{2\sigma_B^2} \right) \times \nonumber \\
&& \exp\left( - \frac{(x-x_i)^2+(y-y_i)^2}{2\sigma^2}\right) . \label{eq:nB-eta}
\end{eqnarray}
The assumption behind this ansatz is that the forward-going participants deposit their baryon charge around space-time rapidity $+\eta_B$, whereas backward-going participants do so at the opposite space-time rapidity $-\eta_B$. Correspondingly, we assume that the local electric charge density $n_Q$ is 0.4 of the local baryon density:
\[ n_Q = 0.4\ n_B. \]
Note that the normalization constant $\kappa_B$ is set so that the total baryon number is conserved in the transition to the fluid dynamical simulation.

3. \textbf{\trento\ IS} introduces a generalized ansatz \cite{Moreland:2014oya} for the entropy density deposition from the participant nucleons as follows:
\begin{equation} T_R(p;T_A,T_B)\equiv\left(\frac{T_A^p+T_B^p}{2}\right)^{1/p} , \end{equation}
where $T_{A,B}$ are the thickness profiles of the two incoming nuclei, and $p$ is a dimensionless parameter which interpolates between the simplified functional forms of the initial entropy density profile in different initial state models, e.g.~$p=1$ corresponds to a Monte Carlo wounded nucleon model, whereas $p=0$ is functionally similar to the notably successful EKRT and IP-Glasma models. \trento\ IS has been successfully used in many studies of $pA$ and $AA$ reactions at the top RHIC and LHC energies. In particular, this initial state model has been used in conjunction with a 2 dimensional event-by-event viscous hydrodynamics + hadronic cascade to constrain the properties of QGP, based on comparison to $\snn=200$~GeV RHIC and $\snn=2.76$ and $5.02$~TeV LHC data employing Bayesian statistics~\cite{Bernhard:2016tnd,Bernhard:2019bmu}. One of the outcomes of the latter studies was a constraint on the $p$ parameter of the \trento\ IS itself. The constrained value is $p=0$, when the reduced thickness is a geometric mean of the thickness profiles from the both nuclei: $T_R=\sqrt{T_A T_B}$.

Since \trento\ IS provides only the transverse density profile, similarly to GLISSANDO IS we assume that the initial entropy density profile in the transverse plane is proportional to the density profile from \trento, and in addition we apply a similar longitudinal structure, described by Eq.~\ref{eq:Heta}, and setting $f_\pm(\eta_s)=1$ as there are no forward- or backward-going sources in the density table from \trento. Likewise, the baryon charge distribution is a version of Eq.~(\ref{eq:nB-eta}) which is symmetric in the space-time rapidity.

\subsection{Total energy and baryon charge counting}
As it is mentioned in the Introduction, the classic 2-dimensional hydrodynamic calculations for the top RHIC or LHC energies rely on the approximation of longitudinal boost invariance. It implies that the hydrodynamic system is infinite in the longitudinal space-time rapidity, and therefore has infinite total energy. As such, the initial energy density for the hydrodynamic stage is essentially a free parameter, which is adjusted in order to reproduce the final charged hadron multiplicity at the mid-rapidity.
At lower collision energies, where the longitudinal boost invariance is not justified anymore, one has to deal with initial energy/entropy density profiles which are finite in all 3 dimensions, and the total initial energy of the hydrodynamic part has to correspond to the initial energy of the participant region.

UrQMD IS accounts for the total energy-momentum of the participant region, because the microscopic processes (string excitation, fragmentation and hadronic scatterings) conserve energy and momentum. Therefore, with the UrQMD IS, the total initial energy of the hydrodynamic system is naturally equal to the total energy of the participant region, $N_\text{W}\snn/2$. The other two initial state models, GLISSANDO and \trento, provide the density distributions in the transverse space, which are then superimposed with the finite longitudinal profiles. Therefore, the total energy of the hydrodynamic system is also finite, but is proportional to the normalization constant $\kappa$ of the initial energy density profile.

In case of GLISSANDO and \trento\ IS, in each initial state configuration in the event-by-event ensemble we determine the values of $\kappa$ and $\kappa_B$ numerically, so that the total energy in the initial state and the total baryon charge are equal to $N_\text{W}\snn/2$ and $N_\text{W}$, respectively:
\begin{eqnarray}
\tau_0 \int \epsilon\cosh\eta\cdot dxdyd\eta & = & \frac{N_\text{W}}{2}\snn \\
\tau_0 \int n_B dxdyd\eta & =& N_\text{W}.
\end{eqnarray}
Obviously, the actual values of $\kappa$ and $\kappa_B$  depend on the choice for $\eta_0$. Larger values of $\eta_0$ put more energy into the longitudinal motion instead of increasing the energy density. However, the rapidity distributions of hadrons are primarily sensitive to this, thus we have an independent observable for the determination of $\eta_0$. Subsequently, $\kappa$'s can be calculated.

Furthermore, the hydrodynamic stage itself conserves the total energy up to numerical errors. Therefore, the procedure above ensures that the total energy of the system is conserved in the pre-hydrodynamic and hydrodynamic stages. The only source of the variation of total energy of the final-state hadrons with respect to the initial energy $N_\text{W}\snn/2$ is the grand-canonical sampling of hadrons at the fluid-to-particle (particlization) hypersurface, as will be discussed in the next subsection.

\subsection{Hydrodynamic and post-hydrodynamic stages}
The hydrodynamic stage of evolution is modelled with a 3 dimensional relativistic viscous hydrodynamic code \texttt{vHLLE}~\cite{Karpenko:2013wva}. The code numerically solves the equations of relativistic viscous fluid dynamics in the Israel-Stewart framework, namely the energy-momentum and baryon number conservation:
\begin{equation} \nabla_{\nu} T^{\mu\nu}=0\nonumber,\quad \nabla_{\nu}N^\nu=0\ , \end{equation}
and the evolution  equations for the shear stress tensor:
\begin{equation}
\left\langle u^\gamma \nabla_{\gamma} \pi^{\mu\nu}\right\rangle =-\frac{\pi^{\mu\nu}-\pi_\text{NS}^{\mu\nu}}{\tau_\pi}-\frac 4 3 \pi^{\mu\nu}\nabla_{\gamma}u^\gamma
\end{equation}
where $\nabla_{\mu}$ denotes the covariant derivative in Milne coordinates, $\pi_\text{NS}^{\mu\nu}$ is the shear stress tensor in Navier-Stokes limit.

For the equation of state (EoS) in the hydrodynamic stage, we follow the choice from~\cite{Karpenko:2015xea} and use the chiral model EoS~\cite{Steinheimer:2010ib}, which comprises correct degrees of freedom, i.e., hadrons at low temperature and quarks and gluons at high temperature limits. The EoS has a crossover-type transition between hadronic and partonic phases for all values of the baryon chemical potential $\mu_B$ and qualitatively agrees with the lattice QCD calculations at $\mu_B=0$.

Likewise, we set fixed, i.e.\ temperature-independent values of the shear viscosity to entropy density ratio $\eta/s$ from~\cite{Karpenko:2015xea}. Bulk viscosity is set to zero, although the hydrodynamic code is capable to evolve the bulk pressure as well. For the relaxation time of the shear stress tensor in the Israel-Stewart equations, an ansatz is made: $\tau_\pi=5\eta/(sT)$.

Fluid-to-particle transition (particlization) takes place at the hyper-surface of fixed energy density $\epsilon_\text{sw}=0.5$~GeV/fm$^3$, when the medium is well in the hadronic phase. The particlization hypersurface is reconstructed in the course of hydrodynamic evolution, timestep-by-timestep and using the Cornelius subroutine~\cite{Huovinen:2012is}. Next, a Monte Carlo hadron sampling is performed according to the phase-space distribution coming from the Cooper-Frye formula~\cite{Cooper:1974mv}. In practice, as the particlization hypersurface is composed of many small elements, the hadron sampling is performed for each element independently (which is consistent with the grand-canonical ensemble) and according to:
\begin{multline}
\frac{d^3 \Delta N_i}{dp^* d({\rm cos}\,\theta)d\phi}=\frac{\Delta\sigma^*_\mu p^{*\mu}}{p^{*0}} 
p^{*2} f_\text{eq}(p^{*0};T,\mu_i) \\
\times\left[ 1+(1\mp f_\text{eq})\frac{p^*_\mu p^*_\nu \pi^{*\mu\nu}}{2T^2(\epsilon+p)} \right]. \label{DF-LRF-visc}
\end{multline}
where the $*$ superscript refers to the quantities in the local fluid rest frame.
The hadrons, sampled at the particlization hypersurface, are then passed on to the UrQMD cascade~\cite{Bass:1998ca} to simulate hadronic rescatterings in the post-hydrodynamic stage, as well as resonance decays.

To enhance the statistics of the generated hadronic events, we apply a so-called oversampling procedure: for each hydrodynamic evolution and the corresponding particlization hypersurface, we repeat the hadron sampling according to the Cooper-Frye few hundreds of times. Such few hundreds of intermediate hadronic events are then passed separately and independently to the UrQMD cascade.


\section{Results}\label{sect-results}
In this work we consider only two energies in the RHIC BES range: $\snn=27$ and 62.4~GeV, and do not extend the simulations to lower energies. Such choice is made since at $\snn=27$~GeV the Lorentz contraction of the colliding nuclei is still relatively strong, such that the average time for the two nuclei to completely pass through each other is not larger than 1~fm/c. As such, the picture of colliding thin disks or pancakes still approximately applies.

\begin{table}
\begin{center}
\begin{tabular}{|c|c|c|}
\hline
   ~   & \multicolumn{2}{|c|}{$\snn$~[GeV]} \\ \hline
   ~   & 27 & 62.4 \\ \hline \hline
     centrality [\%]  &  $N_\text{W}$ & $N_\text{W}$  \\ \hline
     0  &  394  &  394  \\ \hline
     5  &  321  &  327  \\ \hline
     10  &  272  &  274  \\ \hline
     20  &  196  &  197  \\ \hline
     30  &  138  &  139  \\ \hline
     40  &  93  &  95  \\ \hline
     50  &  60  &  62  \\ \hline
 \end{tabular}
\end{center}
\caption{The upper limits on the numbers of wounded nucleons for given centrality percentile, for $\snn=27$ and 62.4~GeV, used throughout the study.}\label{tb:centrality-bins}
\end{table}

\paragraph{Definition of centrality classes.} Across all initial state scenarios, the centrality classes are defined in the same way, as fixed ranges of the number of wounded nucleons $N_\text{W}$. The ranges, shown in Table~\ref{tb:centrality-bins}, are fixed by binning a set of minimum-bias events in terms of the RDS variable defined in the GLISSANDO code, assuming the values of the mixing parameter $\alpha=0.123$ and $0.132$ for $\snn=27$~GeV and $\snn=62.4$~GeV, respectively (see Subsection~\ref{subsect-IS}). The resulting ranges are consistent with the classes defined by the  STAR collaboration, e.g.,~\cite{Adamczyk:2012ku}.

In practice, in each scenario we generate a large set of final hadronic events corresponding to a wide centrality range $5-50\%$. The values of $N_w$ are extracted from the initial state configurations and are recorded in the corresponding final-state events. Whereas GLISSANDO and \trento\ IS provide a direct output of $N_w$, with UrQMD IS we use the total baryon charge of the initial state as a proxy for $N_w$. The centrality selection according to the Table~\ref{tb:centrality-bins} is then applied {\it after} the simulation, by selecting the events with given $N_w$ to compute the observables.

The parameters of the longitudinal structure in GLISSANDO and \trento\ IS are summarized in Table~\ref{tb:long-params}. Whereas for $\snn=62.4$~GeV the parameters do not depend on centrality, adjustment to the experimentally measured $dN_\text{ch}/d\eta$ and $dN_\text{net p}/dy$ suggested a somewhat different amount of baryon stopping at different centralities. Therefore the longitudinal parameters were chosen to slightly change with a centrality measure $\chi=N_\text{W}/(2A)$, where $A=197$ is the mass number of the gold nucleus.

\begin{table*}
\begin{center}
\begin{tabular}{|c|c|c|c|c|c|c|c|c}
\hline
 $\snn$~[GeV] & $\tau_0$~[fm/c] & $\eta_0$ & $\sigma_\eta$ & $\eta_M$ & $\eta_B$ & $\sigma_B$ & $\eta/s$ \\ \hline\hline
     27           &      1.0    &  $0.89-0.2\chi$ & $1.09-0.2\chi$  & 1.8 &  $1.33 - 0.32\chi$  &  $0.79 - 0.21\chi$ &  0.12  \\ \hline
     62.4         &      0.7    &   1.8  &  0.7   &  1.0   &   2.2  &  1.0   &    0.08  \\ \hline
 \end{tabular}
\caption{Default values of the model parameter with GLISSANDO/\trento\ IS: starting time, parameters for the longitudinal profile of GLISSANDO and \trento\ IS for $\snn=27$ and 62.4~GeV. In the $\snn=27$~row, a centrality measure $\chi=N_\text{W}/(2A)$ is introduced.}\label{tb:long-params}
\end{center}
\end{table*}

\begin{figure} \centering
\includegraphics[width=0.5\textwidth]{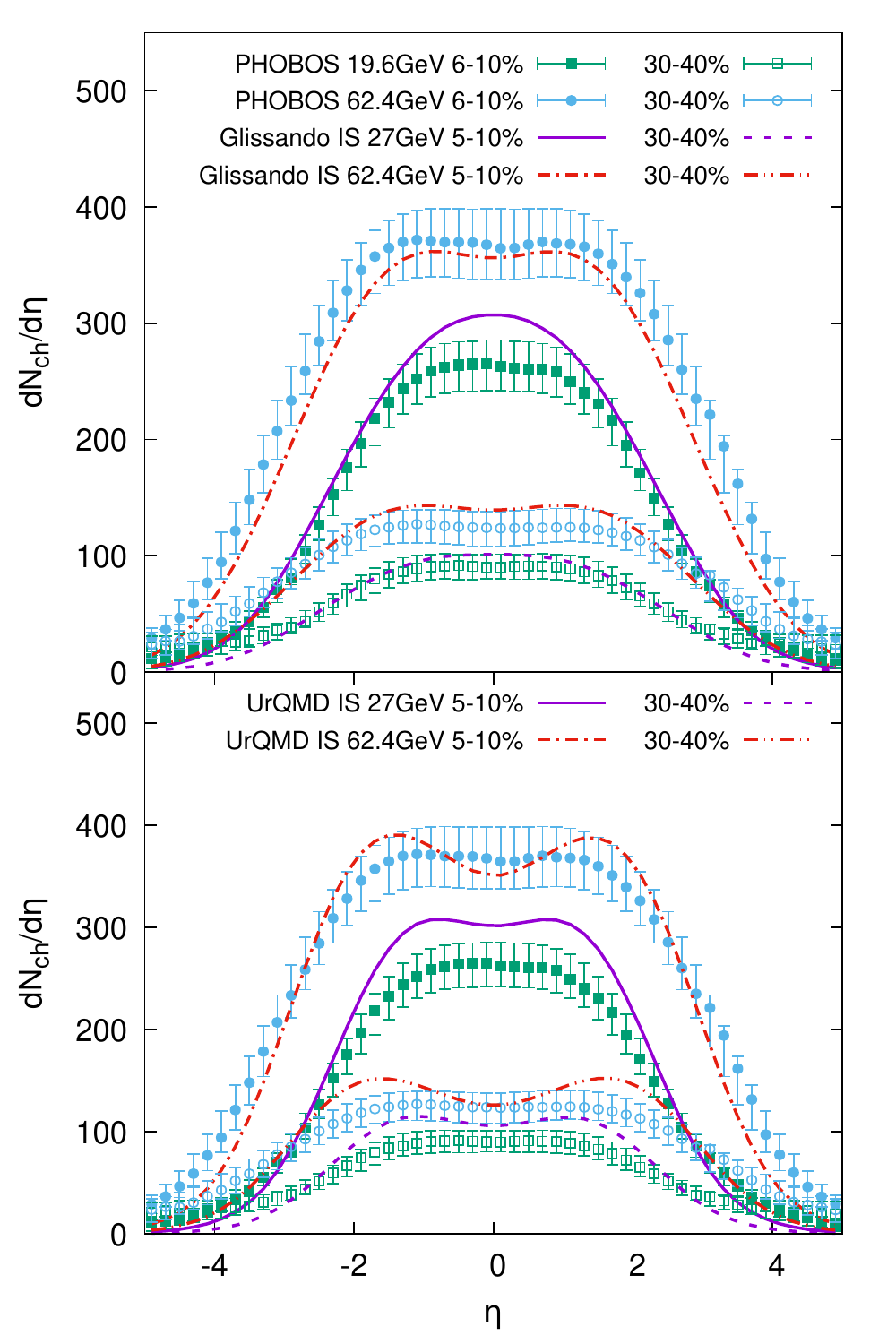}
\caption{Pseudorapidity density of charged hadrons in Au-Au collisions at $\snn=27$ and 62.4~GeV from the vHLLE+UrQMD simulations with GLISSANDO and UrQMD initial states. As no experimental data points are available for $\snn=27$~GeV, the data points from~\cite{Alver:2010ck} are plotted for neighbouring energy $\snn=19.6$~GeV.}\label{fig:dndy}
\end{figure}

\paragraph{Rapidity distributions.} We start with the most basic observable model-wise: the pseudo-rapidity density of charged hadrons. The pseudo-rapidity distributions are shown in Fig.~\ref{fig:dndy} for the UrQMD IS and GLISSANDO IS. \trento\ IS results in the pseudo-rapidity distributions, which are very close to the GLISSANDO IS case, therefore we omit them in the plot in order not to make it too busy. Whereas the $p_T$ spectra and elliptic flow have been measured for many BES energies by STAR, the rapidity distributions were generally not, therefore we use older experimental data from PHOBOS collaboration~\cite{Alver:2010ck} as a reference, taken at $\snn=19.6$ and 62.4~GeV. One can therefore expect that the pseudorapidity distributions at $\snn=27$~GeV from the model calculation should be found between the $\snn=19.6$ and 62.4~GeV points from the data. One can notice that the $dN_\text{ch}/d\eta$ within $|\eta|<2$ is somewhat larger with UrQMD IS as compared to GLISSANDO or \trento\ IS. The three different initial states provide close values of the mean total energy, baryon and electric charges for events in a given centrality class. The higher $dN_\text{ch}/d\eta$ is therefore a consequence of slightly depleted tails of the rapidity profile with UrQMD IS, as compared to GLISSANDO or \trento\ IS. In fact, with the chosen parameters of the longitudinal structure, GLISSANDO and \trento\ IS approach the experimental data points closer at large pseudorapidities. However, the depleted tails of the rapidity profile with UrQMD IS leads to more energy available for the particle production at mid-rapidity, which explains the higher $dN_\text{ch}/d\eta$.

\begin{figure} \centering
\includegraphics[width=0.5\textwidth]{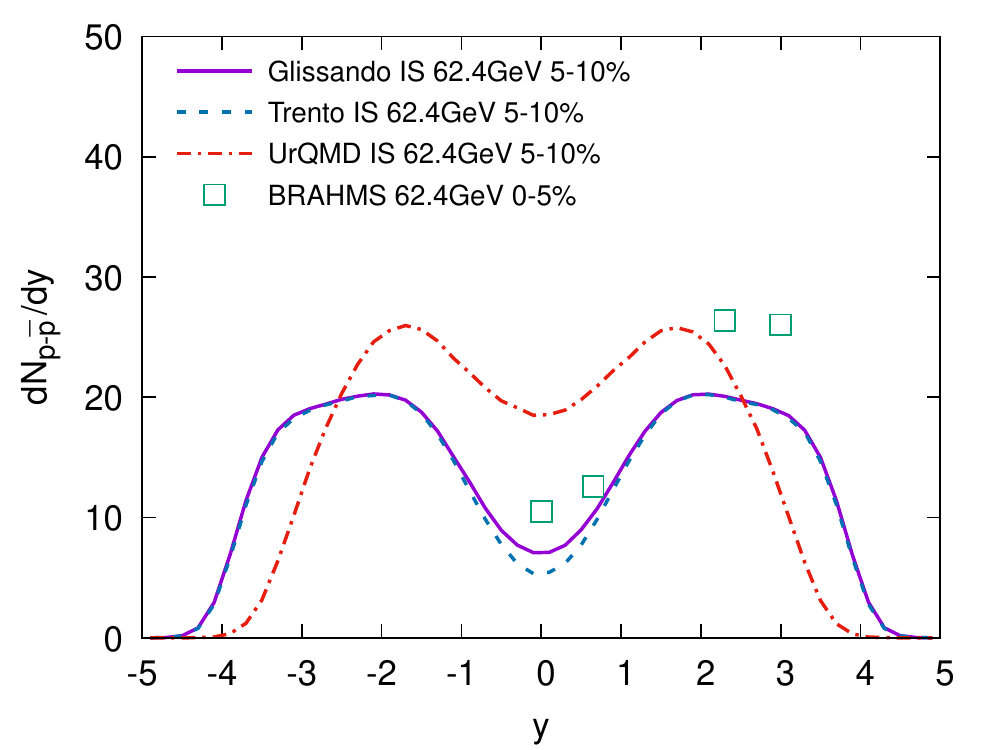}\\
\includegraphics[width=0.5\textwidth]{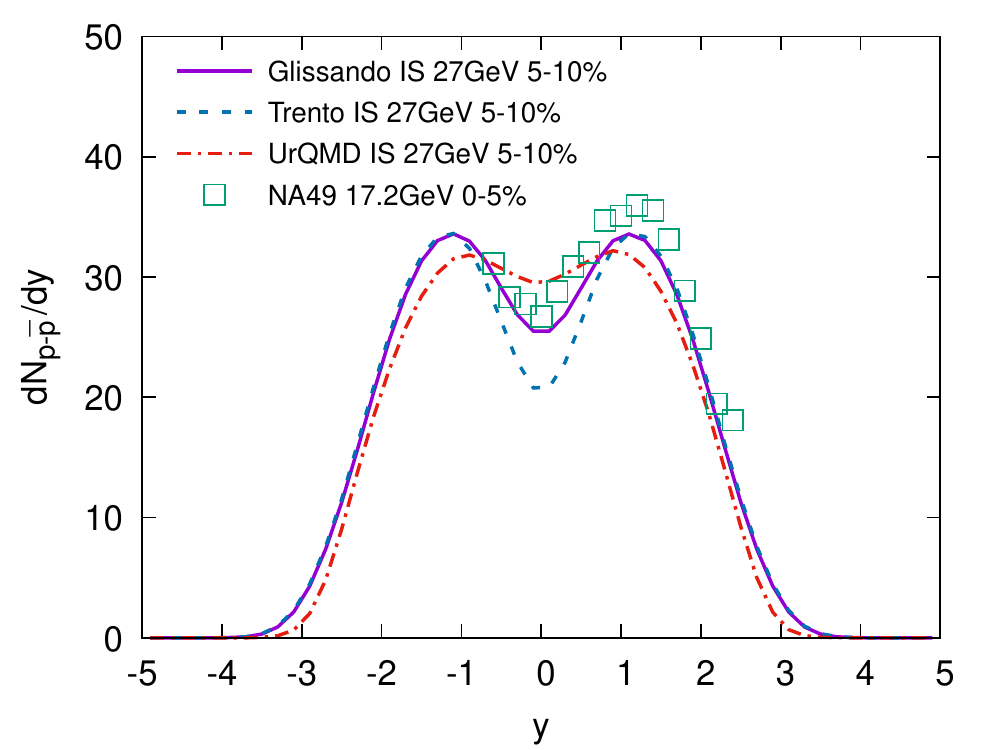}
\caption{Rapidity distribution of net protons at $\snn=62.4$~GeV (top panel) and $\snn=27$~GeV (bottom panel) from the vHLLE+UrQMD simulations with UrQMD, GLISSANDO and \trento\ initial states. The experimental data points for 0-5\% central Au-Au collisions are taken from BRAHMS \cite{Arsene:2009aa}, whereas the model calculations correspond to 5-10\% centrality. The experimental data points for $\snn=17.2$~GeV are taken from NA49 \cite{Appelshauser:1998yb}.}\label{fig:dn-net-protons}
\end{figure}

At the collision energies considered, baryon density becomes non-negligible. Therefore on Fig.~\ref{fig:dn-net-protons} we show the rapidity distribution of net protons (protons minus anti-protons) at $\snn=27$ (bottom panel) and 62.4 GeV (top panel) from the model with different IS scenarios in comparison to experimental data. At $\snn=62.4$ GeV, the experimental data points from BRAHMS~\cite{Arsene:2009aa} are used, whereas the results at $\snn=27$~GeV are compared to the data from Pb-Pb collisions measured by NA49 experiment at $\snn=17.2$~GeV ($E_\text{lab}=158$~GeV). With both GLISSANDO and \trento\ IS, the longitudinal shape of the initial baryon density profile can be adjusted with the parameter $\eta_B$, while the normalization procedure ensures fixed total baryon charge regardless of the $\eta_B$ value. The value of this parameter was set as in Table~\ref{tb:long-params} in order to approach the experimental rapidity distribution of net protons, measured by the BRAHMS collaboration. However, with the UrQMD IS, the initial baryon density profile is fixed, and it leads to a narrower final-state net proton rapidity distribution, which goes above the experimental data points at the mid-rapidity $y\approx 0$.

\begin{figure} \centering
\includegraphics[width=0.5\textwidth]{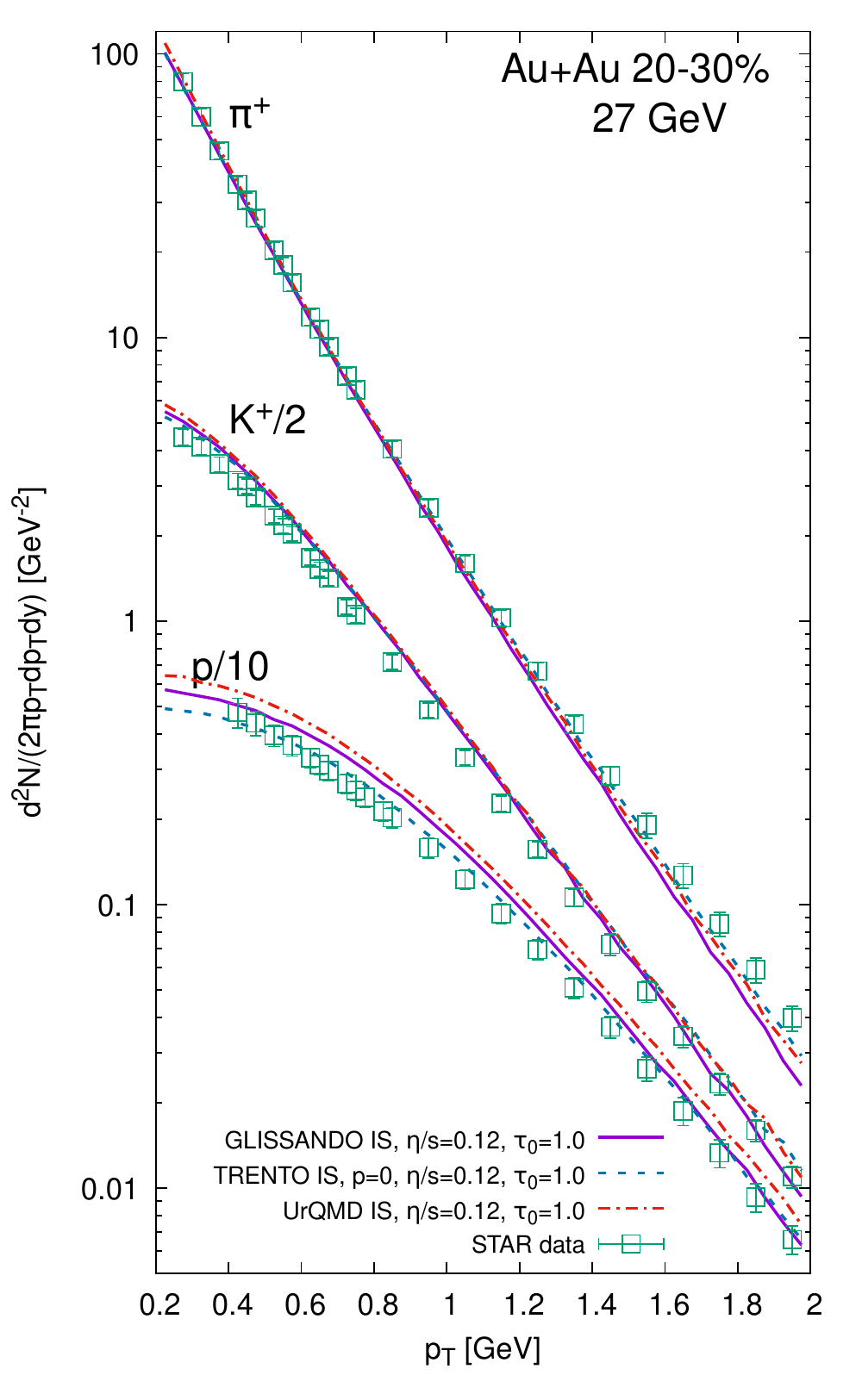}
\caption{Transverse momentum spectra of positively charged pions, kaons and protons at $\snn=27$~GeV from the vHLLE+UrQMD simulations with UrQMD, GLISSANDO and \trento\ initial states. The experimental data points for 20-30\% central Au-Au collisions are taken from \cite{Adamczyk:2017iwn}.}\label{fig:pt-spectra-27gev}
\end{figure}

\begin{figure} \centering
\includegraphics[width=0.5\textwidth]{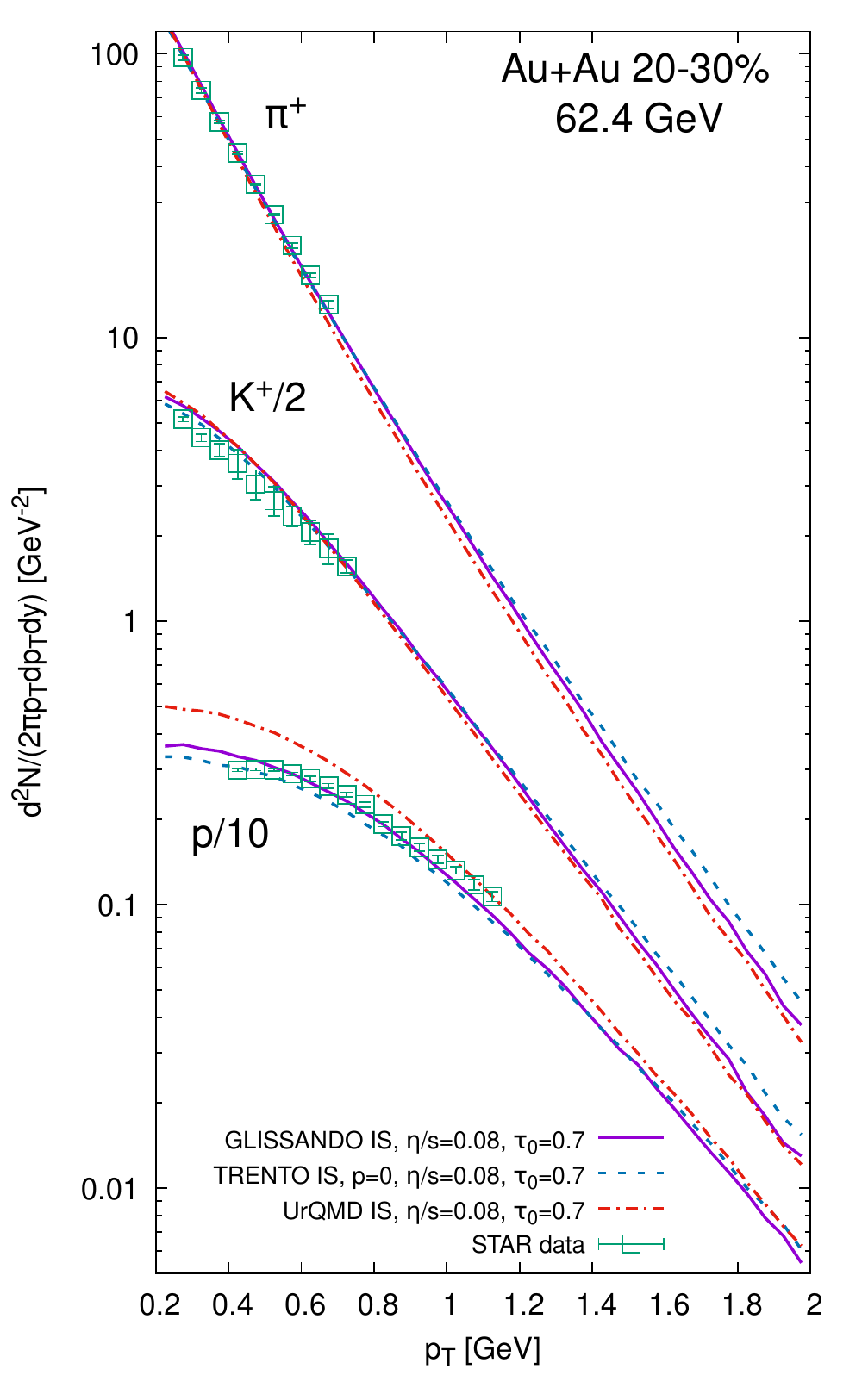}
\caption{Same as Fig.~\ref{fig:pt-spectra-27gev} but for $\snn=62.4$~GeV. The experimental data points for 20-30\% central Au-Au collisions are taken from \cite{Abelev:2008ab}.}\label{fig:pt-spectra-62gev}
\end{figure}

\paragraph{Transverse momentum distributions.} Next, we turn to the transverse momentum ($p_T$) spectra of identified hadrons at mid-rapidity. Fig.~\ref{fig:pt-spectra-27gev} shows the $p_T$ spectra of positively charged pions, kaons and protons for 20-30\% central Au-Au collisions at $\snn=27$~GeV. One can see that, whereas the shape of the kaon $p_T$ spectrum is reproduced with the different initial state models almost equally well, there are small differences in the slopes of pion and kaon spectra between the different IS scenarios, which come from somewhat different strength of the radial flow. The most noticeable difference is in the magnitude of the proton $p_T$ spectrum, which originates in the different baryon charge distribution in rapidity with the different IS. We argue that the \trento\ IS scenario provides the best combined description of all 3 spectra. Fig.~\ref{fig:pt-spectra-62gev} shows the $p_T$ spectra for $\snn=62.4$~GeV collision energy. One can note the same trends with the different scenarios of the initial state, with \trento\ IS resulting in the best combined description of all 3 $p_T$ spectra. However, due to the lack of high-quality experimental data points at this energy, it is not easy make a conclusive judgement.

\begin{figure} \centering
\includegraphics[width=0.5\textwidth]{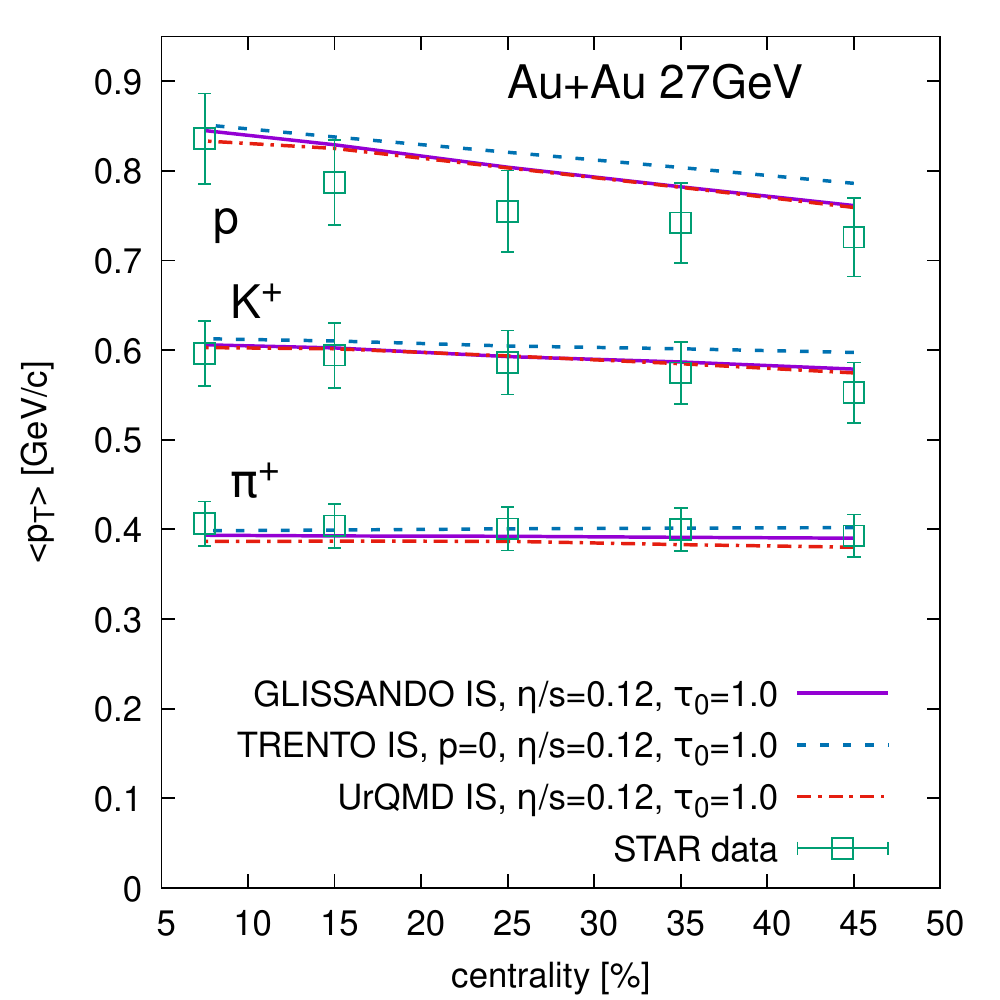}
\caption{Mean transverse momentum of positively charged pions, kaons and protons as a function of centrality at $\snn=27$~GeV from the vHLLE+UrQMD simulations with UrQMD, GLISSANDO and \trento\ initial states. The experimental data points for 20-30\% central Au-Au collisions are taken from \cite{Adamczyk:2017iwn}.}\label{fig:meanpt-27gev}
\end{figure}

\begin{figure} \centering
\includegraphics[width=0.5\textwidth]{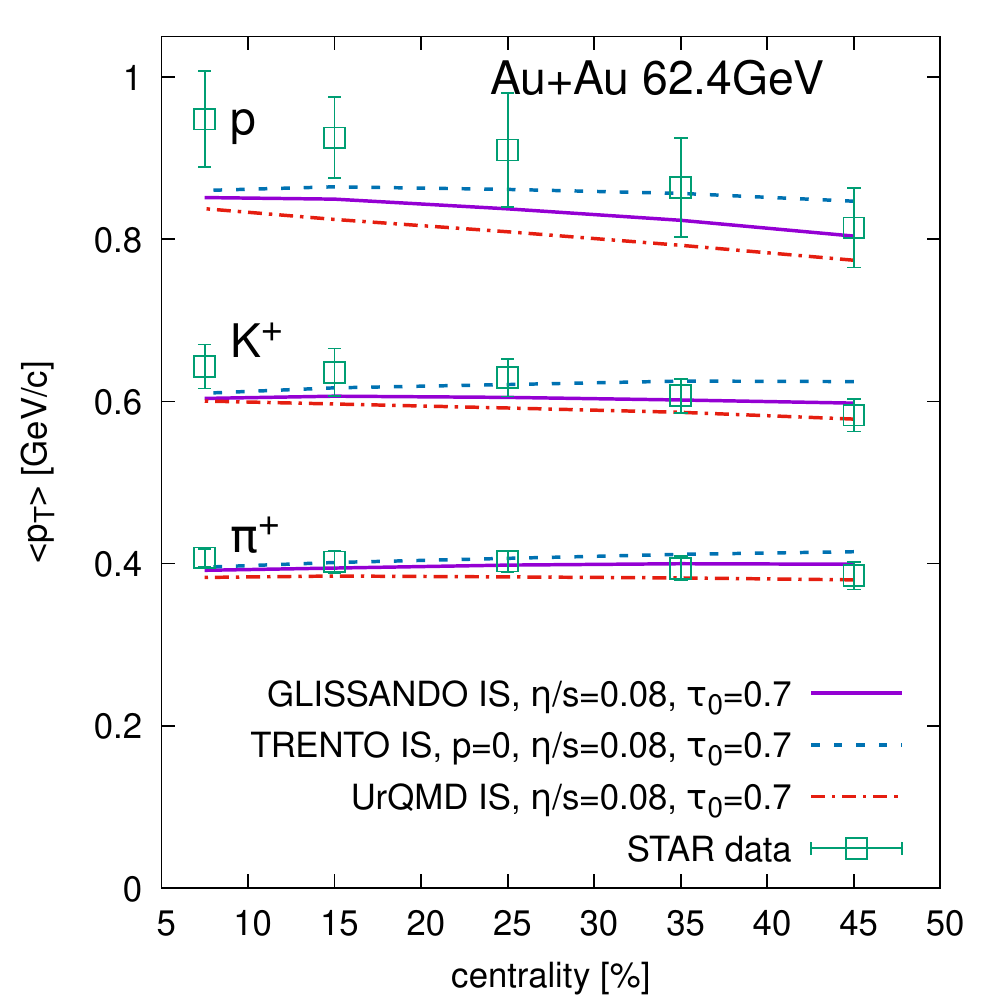}
\caption{Same as Fig.~\ref{fig:meanpt-27gev} but for $\snn=62.4$~GeV. The experimental data points for 20-30\% central Au-Au collisions are taken from \cite{Abelev:2008ab}.}\label{fig:meanpt-62gev}
\end{figure}

To examine the properties of $p_T$ spectra at all centralities it is instructive to look at its quantifiable property, i.e.\ the mean $p_T$. The mean $p_T$ of positively charged pions, kaons and protons as  functions of centrality are shown in Fig.~\ref{fig:meanpt-27gev} for $\snn=27$~GeV and in Fig.~\ref{fig:meanpt-62gev} for $\snn=62.4$~GeV. At $\snn=27$~GeV, the mean $p_T$ of protons is slightly above the data for all of the IS scenarios, whereas the mean $p_T$ of both kaons and pions are within the experimental error bars. This suggests that the radial flow at mid-rapidity is slightly too strong with all three IS scenarios. At $\snn=62.4$~GeV, there is a reverse trend: with UrQMD and GLISSANDO IS, the mean $p_T$ of protons and also kaons to some extent, are below the data points, whereas the mean $p_T$ of pions is consistent with the data, suggesting too weak radial flow at this collision energy. Also, at $\snn=62.4$~GeV, the trends in centrality dependence are somewhat opposite to the data. The mean $p_T$ of kaons and protons show a decreasing trend with centrality in the experiment, whereas in the model the mean $p_T$ of kaons stays flat or increases with GLISSANDO and \trento\ IS respectively, and the mean $p_T$ of protons does not decrease as fast as the data does. The UrQMD IS scenario produces the best centrality dependence for all the species, albeit at this collision energy it underestimates the magnitude of the mean $p_T$ of both protons and kaons. The different centrality trends at $\snn=27$ and $62.4$ GeV with GLISSANDO and \trento\ IS hint that the nuclear stopping should depend on the centrality also at $\snn=62.4$~GeV, via the centrality-dependent parameters of the longitudinal profiles in the initial state.

\begin{figure} \centering
\includegraphics[width=0.5\textwidth]{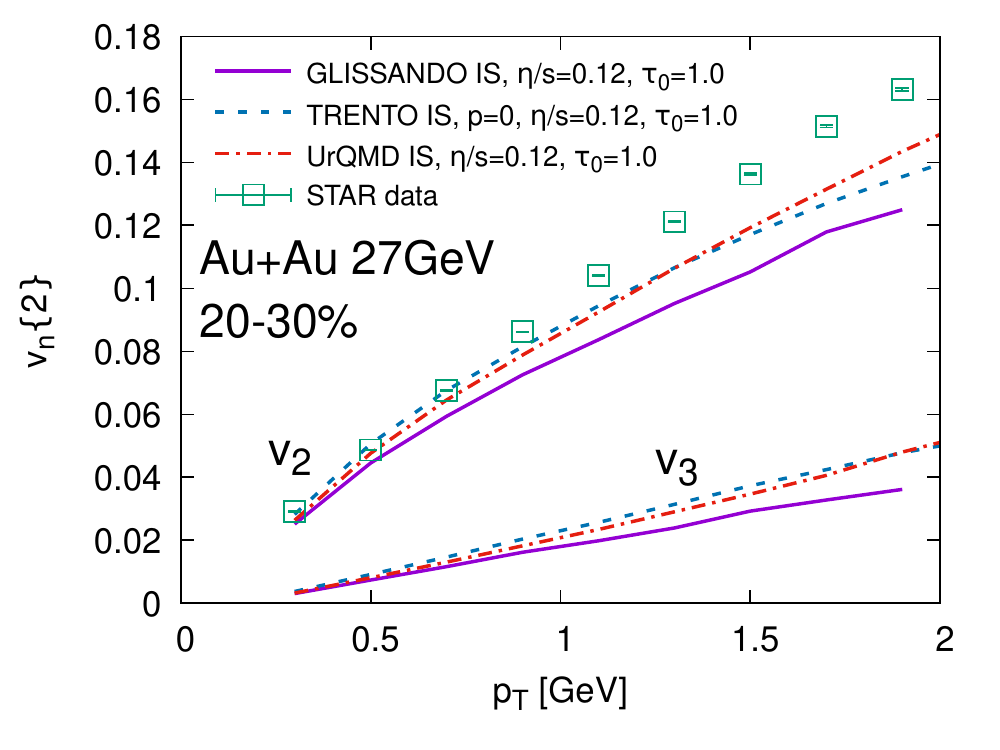}
\caption{Elliptic and triangular flows of all charged hadrons as a function of transverse momentum $p_T$, for 20-30\% Au-Au collisions at $\snn=27$~GeV, from the calculations using UrQMD, Glissando and \trento\ initial state. The experimental data points are taken from \cite{Adamczyk:2012ku}.}\label{fig:v2-pt-27gev}
\end{figure}

\begin{figure} \centering
\includegraphics[width=0.5\textwidth]{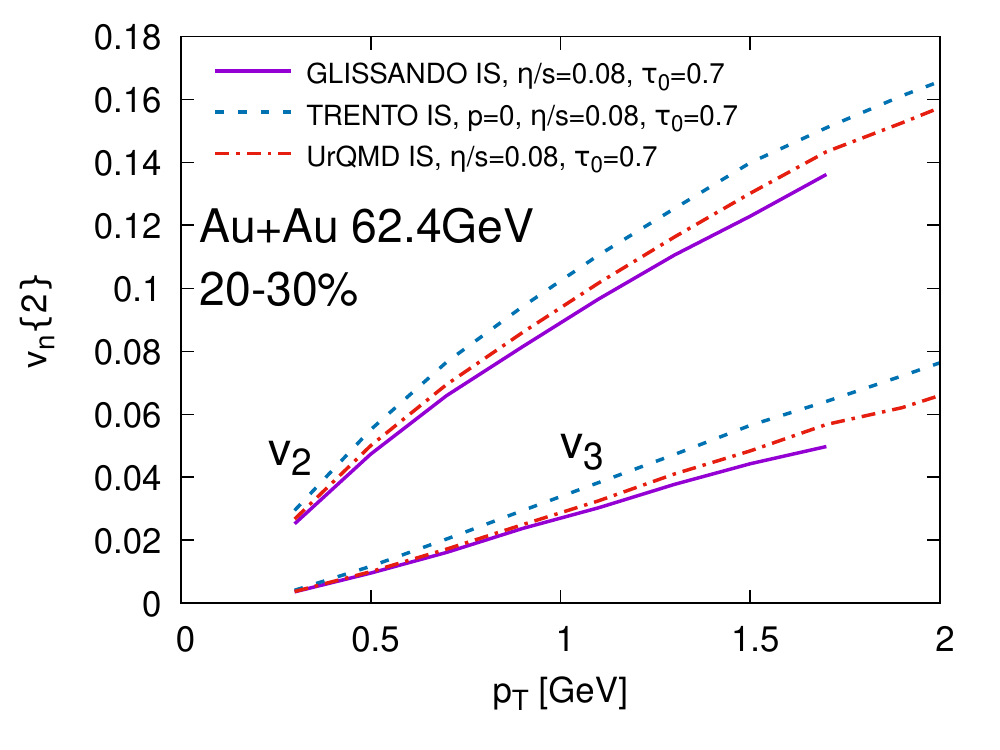}
\caption{Same as Fig.~\ref{fig:v2-pt-27gev} but for $\snn=62$~GeV Au-Au collisions.}\label{fig:v2-pt-62gev}
\end{figure}

\begin{figure} \centering
\includegraphics[width=0.5\textwidth]{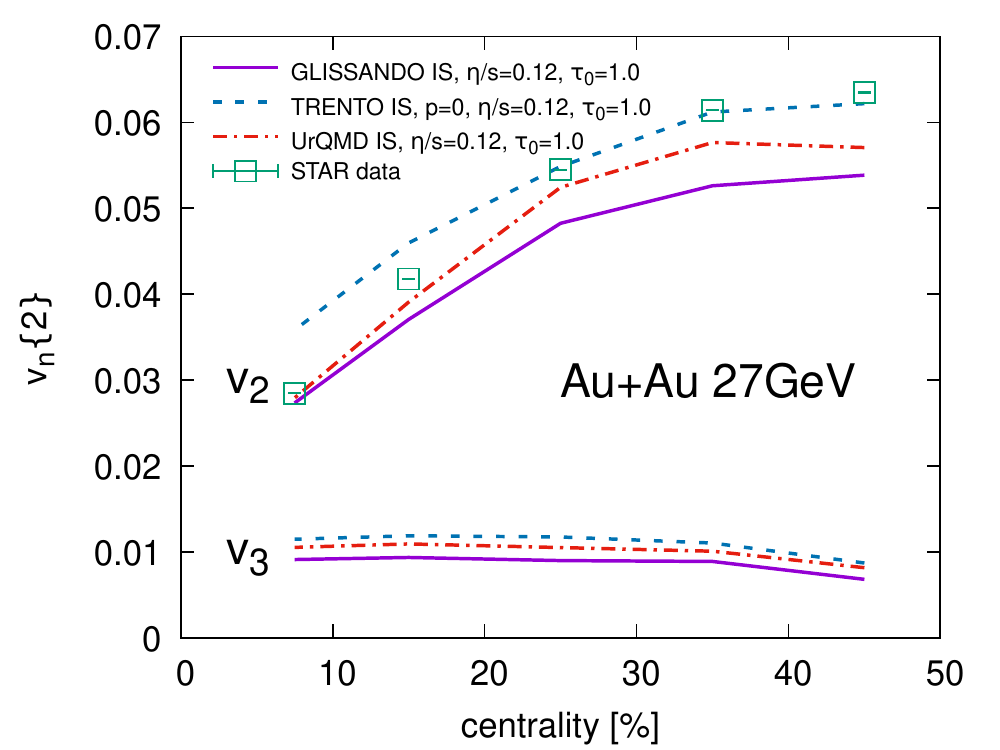}
\caption{Elliptic and triangular flows of all charged hadrons as a function of centrality for $\snn=27$~GeV Au-Au collisions, from the calculations using UrQMD, Glissando and \trento\ initial state. The experimental data points are taken from \cite{Adamczyk:2012ku}.}\label{fig:v2-centr-27gev}
\end{figure}

\begin{figure} \centering
\includegraphics[width=0.5\textwidth]{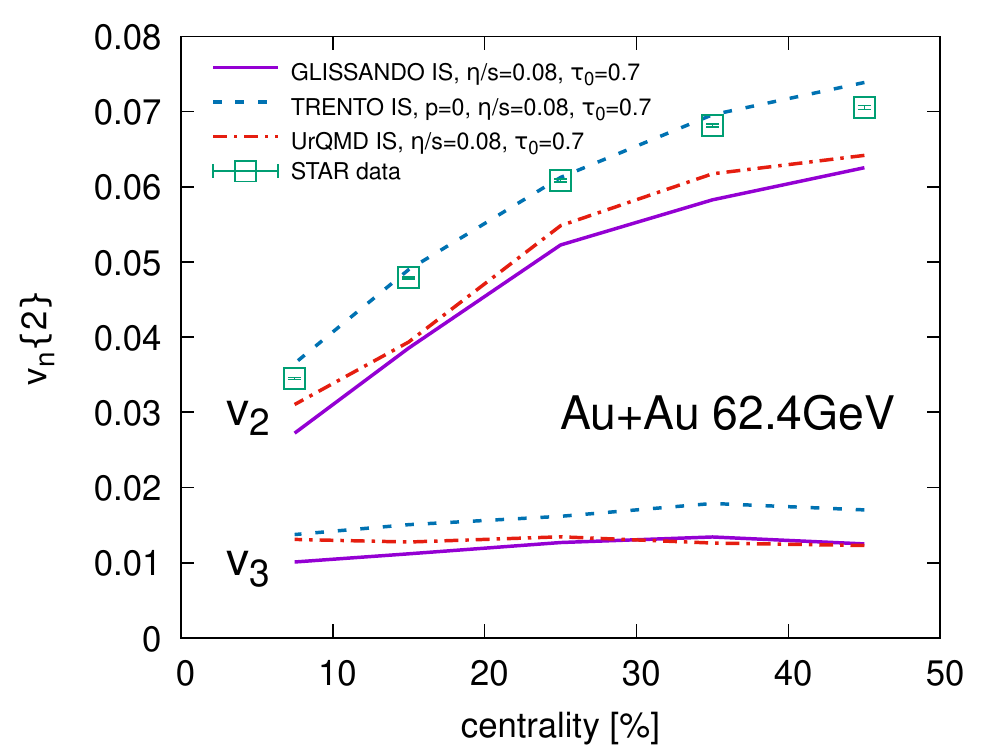}
\caption{Same as Fig.~\ref{fig:v2-centr-27gev} but for $\snn=62$~GeV Au-Au collisions. The experimental data points are taken from \cite{Adamczyk:2017hdl}.}\label{fig:v2-centr-62gev}
\end{figure}

\paragraph{Anisotropic flow.} Finally, we discuss the anisotropic flow observables. Figures~\ref{fig:v2-pt-27gev} and \ref{fig:v2-pt-62gev} show the transverse momentum dependent elliptic flow of all charged hadrons in 20-30\% central Au-Au collisions at $\snn=27$ and 62.4~GeV respectively. The elliptic flow has been computed using the 2-particle cumulant method, following the procedure described in~\cite{Bilandzic:2010jr}. The cumulant method typically requires  rather large event statistics to achieve a reasonably small statistical error. Matching the experimental event statistics in a hydrodynamic model calculation is a CPU-intensive task, therefore we employ a technique which we call ``super-events''. Namely, the final-state events coming from the same underlying hydrodynamic evolution (via the oversampling technique, see above) are considered as one super-event for the purpose of computing the $n$-particle cumulants.

From Fig.~\ref{fig:v2-pt-27gev} one can see that at $p_T<1$~GeV the calculations with \trento\ and UrQMD IS are in agreement with the $v_2\{2\}$ data from STAR, whereas at $p_T>1$~GeV, with the standard settings for the hydrodynamic stage, $\eta/s=0.12$ and $\tau_0=1$~fm/c, all of the calculations start to under-predict the data. Since the $p_T$ spectrum of charged hadrons is steeply falling with $p_T$, the agreement with the data at low $p_T$ is particularly important to reproduce the $p_T$-integrated elliptic flow, as it will be seen later. A similar comparison to the experimental data points at $\snn=62.4$~GeV is not possible due to their absence. The $p_T$ dependence of the triangular flow $v_3\{2\}$ shows the same hierarchy between the different initial state models. Due to the absence of the experimental data points for $v_2$ and $v_3$ we leave the model results as predictions (Fig.~\ref{fig:v2-pt-62gev}), which may be confronted in the proposed \AFTER \ experiment.

In Figs.~\ref{fig:v2-centr-27gev} and \ref{fig:v2-centr-62gev} we show the centrality dependence of the $p_T$-integrated elliptic flow for Au-Au collisions at $\snn=27$ and 62.4~GeV, respectively. In the $p_T$-integrated flow, we observe the same hierarchy between the calculations with different initial states: \trento\ IS with $p=0$ produces the largest elliptic and triangular flow, whereas GLISSANDO IS produces the lowest $v_2$ and $v_3$. In the non-central collisions at $\snn=27$~GeV, as well as in all considered centrality classes at $\snn=62.4$~GeV, calculations with \trento\ IS (p=0) have the best agreement with the experimental data for $v_2\{2\}$.

\begin{figure} \centering
\includegraphics[width=0.5\textwidth]{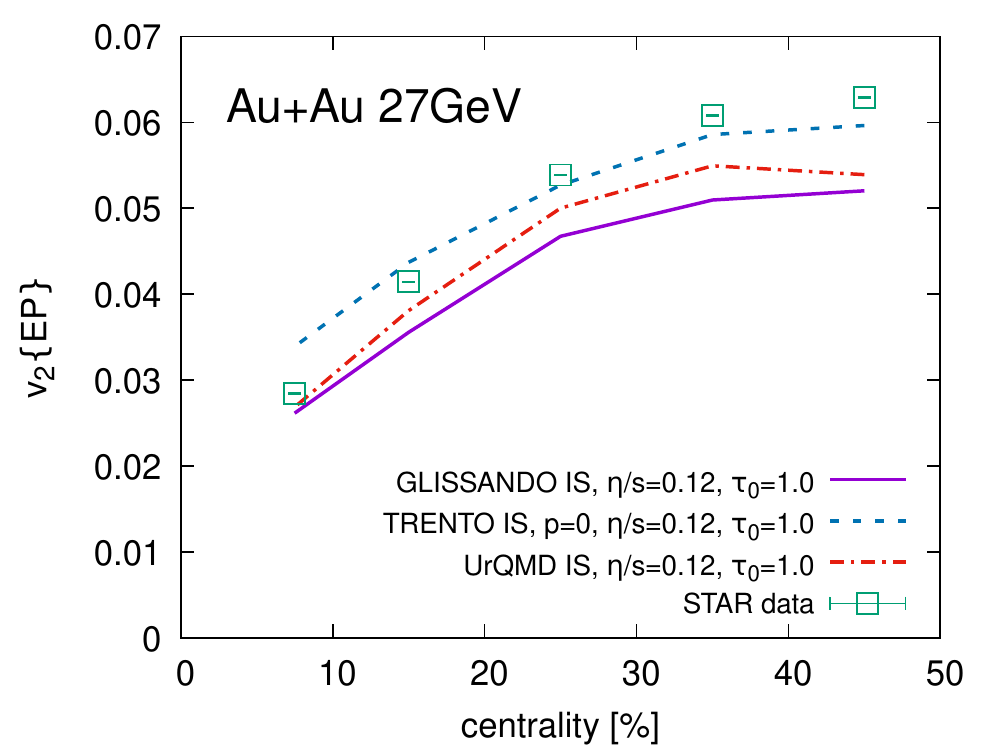}
\caption{Elliptic flow of all charged hadrons as a function of centrality for $\snn=27$~GeV Au-Au collisions, computed with event plane method. The curves represent model calculations using UrQMD, GLISSANDO and \trento\ initial states. The experimental data points are taken from~\cite{Adamczyk:2012ku}.}\label{fig:v2-centr-27gev-EP}
\end{figure}

At this point, we note that in the earlier calculations~\cite{Karpenko:2015xea} with the same UrQMD IS, the $p_T$-integrated elliptic flow in 20-30\% Au-Au collisions at $\snn=27$ and 39~GeV was agreeing with the STAR data rather well. Since the elliptic flow in~\cite{Karpenko:2015xea} was computed with the event-plane method and was compared to $v_2\{{\rm EP}\}$ from STAR, we recreate the same comparison here. For that, we have computed the $p_T$-integrated elliptic flow with the event plane method following the procedure from~\cite{Holopainen:2010gz}, and the result is shown in Fig.~\ref{fig:v2-centr-27gev-EP}. With the event plane method, the elliptic flow in the present calculations shows the same hierarchy between the different initial state models and the same level of agreement with $v_2\{{\rm EP}\}$ from STAR. The actual difference between the present and the earlier calculations lies in the centrality determination. In~\cite{Karpenko:2015xea}, the centrality classes were defined as impact parameter intervals based on a wounded nucleon model, or Glauber model using the public code~\cite{Miskowiec}. In the present calculations, the centrality classes are defined as intervals in the number of wounded nucleons $N_\text{W}$, based on the $N_\text{W}$ binning of minimum-bias nucleus-nucleus collisions with the GLISSANDO 2 code. As such, with the Monte Carlo Glauber approach the mid-central centrality classes correspond to larger values of the average $N_\text{W}$ as compared to~\cite{Miskowiec}. As a consequence, in the present calculation the mid-central events have a slightly lower average initial eccentricity as compared to~\cite{Karpenko:2015xea}.

\begin{figure} \centering
\includegraphics[width=0.5\textwidth]{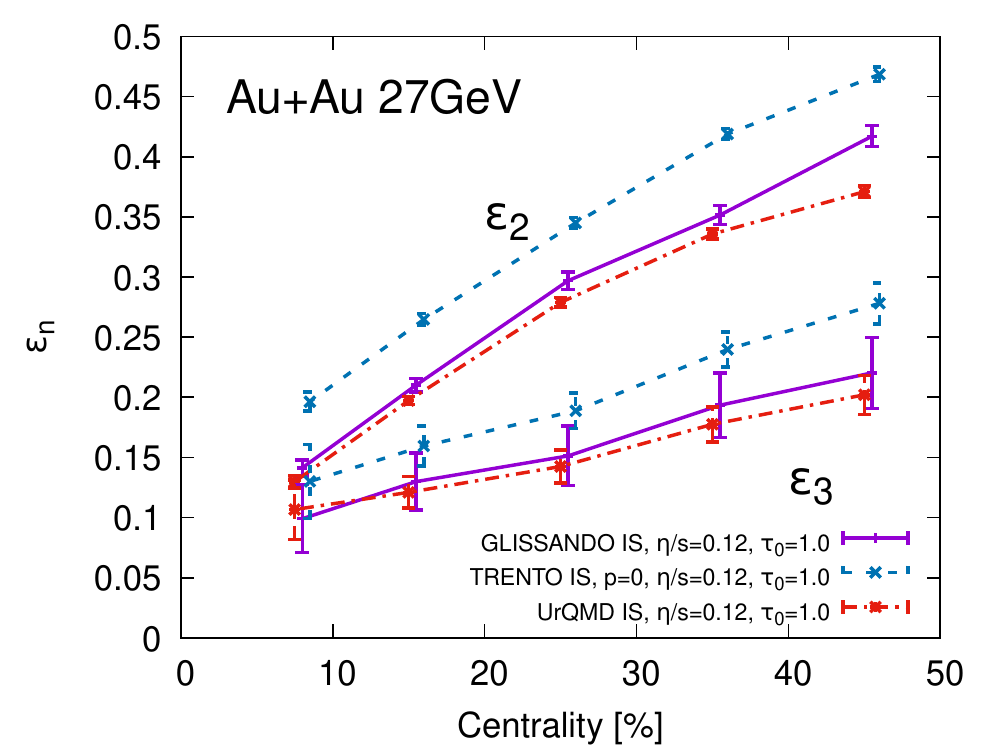}
\caption{Average eccentricities of the initial state energy density as a function of centrality percentile for $\snn=27$~GeV Au-Au collisions. The eccentricities are computed in the UrQMD, GLISSANDO and \trento\ ($p=0$) initial states.}\label{fig:eps2eps3-27}
\end{figure}

\begin{figure} \centering
\includegraphics[width=0.5\textwidth]{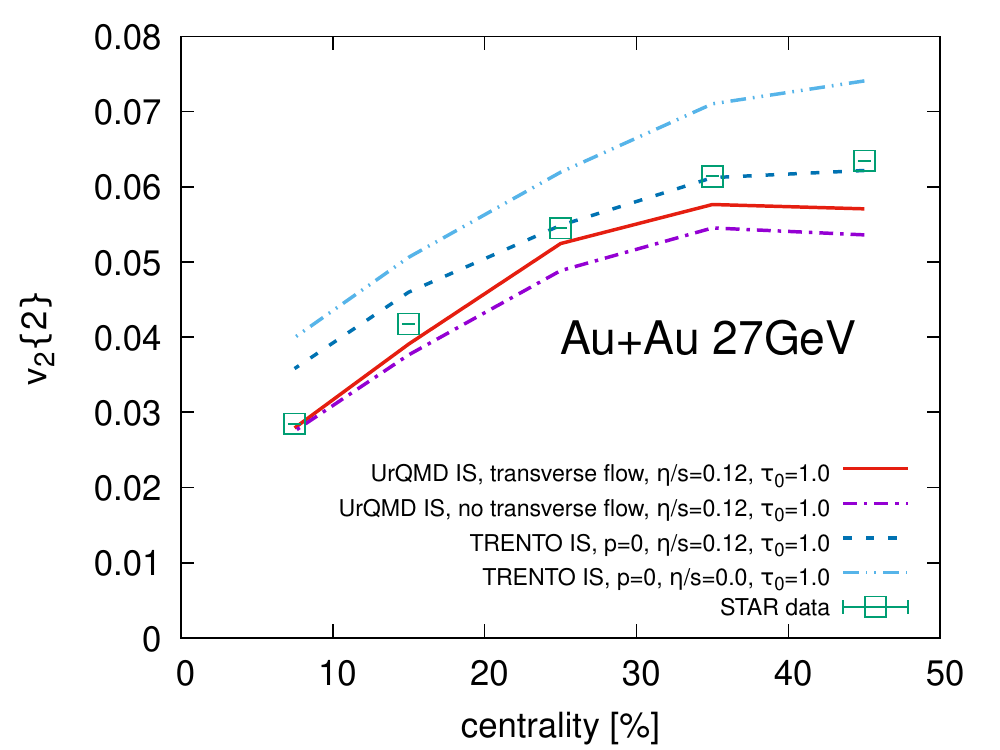}
\caption{Elliptic flow of all charged hadrons as a function of centrality for $\snn=27$~GeV Au-Au collisions, computed with event plane method. The curves represent model calculations using UrQMD initial state with and without transverse flow at the fluidization, and \trento\ initial state with and without the shear viscosity at the fluid stage. The experimental data points are taken from~\cite{Adamczyk:2012ku}.}\label{fig:v2-centr-27gev-noiniflow}
\end{figure}

\begin{figure} \centering
\includegraphics[width=0.5\textwidth]{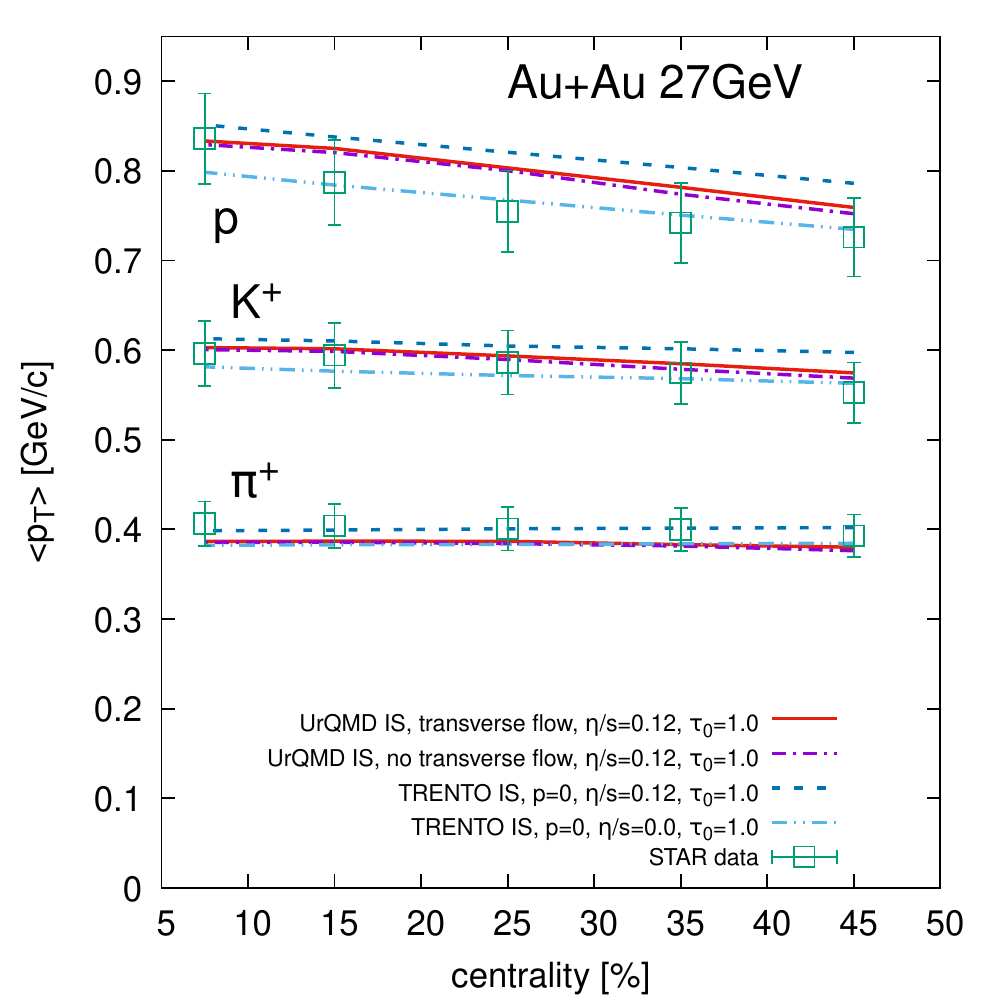}
\caption{Mean transverse momentum of positively charged pions, kaons and protons as a function of centrality for $\snn=27$~GeV Au-Au collisions. The curves represent model calculations using UrQMD initial state with and without transverse flow at the fluidization, and \trento\ initial state with and without the shear viscosity at the fluid stage. The experimental data points are taken from \cite{Adamczyk:2017iwn}.}\label{fig:meanpt-27gev-noiniflow}
\end{figure}

\paragraph{Importance of pre-thermal collective flow and shear viscosity.}
Typically, one assumes that the hierarchy of the elliptic flow from the different initial state originates from the different average eccentricities of the initial energy density profile at the start of the hydrodynamic expansion. To some extent it is true in the present calculations, as can be seen from Fig.~\ref{fig:eps2eps3-27}. However, on this figure one can see that GLISSANDO IS actually has slightly larger average eccentricity and average initial triangularity as compared to UrQMD IS. The reason behind the larger final-state elliptic flow with UrQMD IS is the pre-thermal dynamics in the latter. In UrQMD, the dynamics, and so the development of transverse motion of hadrons, starts with the first nucleon-nucleon scattering and lasts until the hyper-surface of the fluidization $\tau=\sqrt{t^2-z^2}=\tau_0$. In the fluidization procedure in vHLLE, the full 4-momentum of the hadrons crossing the hypersurface of $\tau=\tau_0$ is translated into the 4-velocity of the corresponding fluid cell, which includes the transverse flow velocity. Whereas, both GLISSANDO and \trento\ IS do not incorporate any dynamics, therefore resulting in zero transverse flow at the start of the hydrodynamic stage. To verify the claim, we have performed another set of simulations with UrQMD IS, where the initial transverse flow velocity was re-set to 0, which creates some mismatch between the total energy of the initial state hadrons and the total energy of the fluid. The resulting elliptic flow is shown as a dash-dotted curve in Fig.~\ref{fig:v2-centr-27gev-noiniflow}. One can see that without the transverse flow development at the pre-hydro stage, the elliptic flow with UrQMD IS drops approximately to the level of that with GLISSANDO IS. On the other hand, the pre-hydro flow in UrQMD IS does not visibly affect the mean transverse momentum of hadrons as one can see from Fig.~\ref{fig:meanpt-27gev-noiniflow}.

Another important ingredient for the reproduction of experimental data is the ratio of the shear viscosity to entropy density $\eta/s$ of the fluid medium. One may assume that at the lower collision energies, such as $\snn=27$~GeV, the average duration of the hydrodynamic stage is shorter as compared to the top RHIC or LHC energies, due to the lower initial energy density at the start of it, therefore the details of fluid dynamic modelling (such as the value of $\eta/s$) become less influential. As can be seen from Fig.~\ref{fig:v2-centr-27gev-noiniflow}, the shear viscosity of the medium has a noticeable impact on the final-state elliptic flow also at $\snn=27$~GeV. One can see that the calculation with GLISSANDO IS and zero shear viscosity overshoots the experimental $v_2\{2\}$ in non-central events by 15-20\%. At the same time, in the absence of shear viscous corrections the mean $p_T$ of protons closes in on the experimental data points, as seen in Fig.~\ref{fig:meanpt-27gev-noiniflow} .

Overall, the hybrid model calculation with \trento\ IS provides the best description of the experimental data at $\snn=62.4$~GeV. At this energy, both UrQMD and GLISSANDO IS scenarios result in under-estimated elliptic flow, whereas UrQMD IS case also underestimates the radial flow (as it is seen from the mean $p_T$). The lower elliptic flow with UrQMD and GLISSANDO IS is unlikely to be corrected by a smaller (or zero) value of shear viscosity in the fluid stage, because the latter will drive the mean $p_T$ observable away from the experimental data points. We believe that, similarly to the $\snn=27$~GeV case, the pre-requisite for the elliptic flow is higher initial state eccentricity from \trento\ IS as compared to UrQMD and GLISSANDO IS. This is no surprise as $p=0$ in \trento\ IS results in the initial density profile which is steeper and more spiky than the one from a Monte Carlo Glauber model. Also, the $p=0$ value has been constrained from the experimental transverse momentum spectrum of pions, kaons and protons, yields and flow harmonics $v_2, \dots v_4$ at $\snn=2.76$~TeV LHC energy using Bayesian analysis \cite{Bernhard:2016tnd,Bernhard:2019bmu}.

At $\snn=27$~GeV, there is no clear favourite IS. Whereas the calculation with GLISSANDO IS again tends to underestimate the experimentally measured elliptic flow, this may be fixed by setting a smaller $\eta/s$ in the hydro stage, while staying consistent with the mean $p_T$ observable. \trento\ $p=0$ IS seems to work rather well also at this collision energy. This finding is consistent with the results from~\cite{Shen:2020jwv} which demonstrates that, a 3-dimensional initial state with transverse density profile in a form of $\sqrt{T_A T_B}$, which is the same as \trento\ $p=0$ IS, results in a good agreement with the mean $p_T$ and elliptic flow of hadrons not only at $\snn=27$~GeV but also at the rest of RHIC BES energies. That is an interesting finding, given that the $p=0$ case in \trento\ IS functionally correspond to the notably successful EKRT \cite{Niemi:2015qia} and IP-Glasma~\cite{Schenke:2012wb} models, which are constructed for, and are mostly applicable at higher energies, such as at the LHC.

The beam energy $\snn=62.4$~GeV is considered as the upper end of the Beam Energy Scan program at STAR. Both collision energies considered in this study will not be reachable at the GSI FAIR or JINR NICA facilities. However, one promising future application is the \AFTER\ experiment, which is a proposal of a future fixed-target experiment utilizing the LHCb or ALICE detector at LHC~\cite{Brodsky:2012vg,AFTER,Hadjidakis:2018ifr}. The experiment will collide 2.76~A~TeV lead ion beams on different targets at $\snn=72$~GeV and thanks to the large boost of 4.2 units, it will give an access to a backward center-of-mass rapidity region. The AFTER energy is close to the $\snn=62.4$~GeV considered in this study, therefore adaptation of the model to this energy is quite straightforward. One of the focuses of the AFTER project is the longitudinal structure of the produced QGP medium, which will greatly complement the physics programs of the current collider experiments. It will also test and provide further constraints on the model in the unexplored rapidity domain.


\section{Conclusions}
In this work, we have adapted the initial states from Monte Carlo Glauber (via GLISSANDO 2 code) and $\sqrt{T_A T_B}$ ansatz (via \trento\ code with $p=0$ setting), extended into the longitudinal (space-time rapidity) direction \`{a}-la~\cite{Bozek:2012fw,Bozek:2015bha}, to collision energies $\snn=27$ and $62.4$~GeV. The initial baryon density profiles are introduced along with the initial energy density from GLISSANDO and \trento\ IS, in order to reproduce the rapidity distribution of net protons measured by BRAHMS experiment at RHIC, and by NA49 experiment at the SPS. The total energy and baryon charge in the initial state, and subsequently in the hydrodynamic evolution, are fixed to the total energy and baryon charge of the participant nucleons in a given event. For the latter, the normalization constants of the initial energy density and baryon density profiles are numerically computed, thereby the magnitudes of the initial energy (and also baryon) density are not free parameters as opposed to the classic 2-dimensional fluid dynamic simulations with longitudinal boost invariance.

Both the 3D GLISSANDO and 3D \trento\ ($p=0$) initial states, along with the UrQMD initial state, are coupled to a 3D event-by-event viscous fluid dynamic + cascade model, where the final-state hadronic interactions are simulated with the UrQMD cascade. The fixed values of shear viscosity over entropy density and hydrodynamization time ($\eta/s=0.12, \tau_0=1$~fm/c) and ($\eta/s=0.08, \tau_0=0.7$~fm/c) are used for $\snn=27$ and $62.4$~GeV, respectively, in line with the previous study \cite{Karpenko:2015xea}. We find that both 3D GLISSANDO and 3D \trento\ IS result in an overall fair reproduction of basic experimental data: pseudorapidity distributions, transverse momentum spectra and elliptic flow, at both collision energies. Most notable deviation is a systematic under-estimate of the elliptic flow with GLISSANDO IS, which is rooted in a smaller initial-state eccentricity, as compared to \trento\ IS, and in an absence of transverse dynamics before $\tau=\tau_0$. We note that, somewhat surprisingly, the 3D \trento\ $p=0$ IS, which transverse density profile is functionally similar to the EKRT and IP-Glasma initial state models which work quite well at the LHC energies, describes the basic experimental data reasonably well also at $\snn=27$ and $62.4$~GeV. The latter finding is in line with a recent result from \cite{Shen:2020jwv}, where another version of longitudinally extended initial state based on the $\sqrt{T_A T_B}$ ansatz has been used for the RHIC BES energies $\snn=7.7\dots200$~GeV. Note that, whereas in \cite{Shen:2020jwv} the averaged initial state is used, in this study we run event-by-event hydrodynamics with the fluctuating initial state.

This was rather exploratory study with the aim to map the possibilities and shortcomings of different IS models for the use in hybrid simulations of heavy-ion collisions in the RHIC BES energy region. We would expect that a thorough comparison to data with the use of Bayesian analysis technique \cite{Bernhard:2016tnd} might help to distinguish better between the different models employed here. 

\section{Acknowledgements}
JC, IK, and BT acknowledge support by the project Centre of Advanced Applied Sciences with number CZ.02.1.01/0.0/0.0/16-019/0000778, which is co-financed by the European Union. IK acknwowledges support by the Ministry of Education, Youth and Sports of the Czech Republic under grant ``International Mobility of Researchers – MSCA IF IV at CTU in Prague'' No.\ CZ.02.2.69/0.0/0.0/20\_079/0017983. JC and BAT acknowledge support from from The Czech Science Foundation, grant number: GJ20-16256Y. BT acknowledges support from VEGA 1/0348/18. Computational resources were supplied by the project ``e-Infrastruktura CZ'' (e-INFRA LM2018140) provided within the program Projects of Large Research, Development and Innovations Infrastructures.

\bibliographystyle{h-physrev.bst}
\bibliography{main}

\begin{thebibliography}{10}

\bibitem{Kharzeev:2001yq}
D.~Kharzeev, E.~Levin, and M.~Nardi,
\newblock Phys. Rev. C {\bf 71}, 054903 (2005), hep-ph/0111315.

\bibitem{Schenke:2012wb}
B.~Schenke, P.~Tribedy, and R.~Venugopalan,
\newblock Phys. Rev. Lett. {\bf 108}, 252301 (2012), 1202.6646.

\bibitem{Bernhard:2016tnd}
J.~E. Bernhard, J.~S. Moreland, S.~A. Bass, J.~Liu, and U.~Heinz,
\newblock Phys. Rev. C {\bf 94}, 024907 (2016), 1605.03954.

\bibitem{Bernhard:2019bmu}
J.~E. Bernhard, J.~S. Moreland, and S.~A. Bass,
\newblock Nature Phys. {\bf 15}, 1113 (2019).

\bibitem{Niemi:2015qia}
H.~Niemi, K.~Eskola, and R.~Paatelainen,
\newblock Phys. Rev. C {\bf 93}, 024907 (2016), 1505.02677.

\bibitem{Giacalone:2017uqx}
G.~Giacalone, J.~Noronha-Hostler, and J.-Y. Ollitrault,
\newblock Phys. Rev. C {\bf 95}, 054910 (2017), 1702.01730.

\bibitem{Karpenko:2015xea}
I.~Karpenko, P.~Huovinen, H.~Petersen, and M.~Bleicher,
\newblock Phys. Rev. C {\bf 91}, 064901 (2015), 1502.01978.

\bibitem{Bass:1998ca}
S.~Bass {\em et~al.},
\newblock Prog. Part. Nucl. Phys. {\bf 41}, 255 (1998), nucl-th/9803035.

\bibitem{Ivanov:2005yw}
Y.~Ivanov, V.~Russkikh, and V.~Toneev,
\newblock Phys. Rev. C {\bf 73}, 044904 (2006), nucl-th/0503088.

\bibitem{Weil:2016zrk}
J.~Weil {\em et~al.},
\newblock Phys. Rev. C {\bf 94}, 054905 (2016), 1606.06642.

\bibitem{Rybczynski:2013yba}
M.~Rybczynski, G.~Stefanek, W.~Broniowski, and P.~Bozek,
\newblock Comput. Phys. Commun. {\bf 185}, 1759 (2014), 1310.5475.

\bibitem{Moreland:2014oya}
J.~S. Moreland, J.~E. Bernhard, and S.~A. Bass,
\newblock Phys. Rev. C {\bf 92}, 011901 (2015), 1412.4708.

\bibitem{Karpenko:2013wva}
I.~Karpenko, P.~Huovinen, and M.~Bleicher,
\newblock Comput. Phys. Commun. {\bf 185}, 3016 (2014), 1312.4160.

\bibitem{Cooper:1974mv}
F.~Cooper and G.~Frye,
\newblock Phys. Rev. D {\bf 10}, 186 (1974).

\bibitem{Kharzeev:2000ph}
D.~Kharzeev and M.~Nardi,
\newblock Phys. Lett. B {\bf 507}, 121 (2001), nucl-th/0012025.

\bibitem{Bozek:2012fw}
P.~Bozek and W.~Broniowski,
\newblock Phys. Rev. C {\bf 85}, 044910 (2012), 1203.1810.

\bibitem{Bozek:2015bha}
P.~Bozek, W.~Broniowski, and A.~Olszewski,
\newblock Phys. Rev. C {\bf 91}, 054912 (2015), 1503.07425.

\bibitem{Chatterjee:2017mhc}
S.~Chatterjee and P.~Bozek,
\newblock Phys. Rev. C {\bf 96}, 014906 (2017), 1704.02777.

\bibitem{Bozek:2011if}
P.~Bozek,
\newblock Phys. Rev. C {\bf 85}, 014911 (2012), 1112.0915.

\bibitem{Bozek:2017qir}
P.~Bozek and W.~Broniowski,
\newblock Phys. Rev. C {\bf 97}, 034913 (2018), 1711.03325.

\bibitem{Steinheimer:2010ib}
J.~Steinheimer, S.~Schramm, and H.~Stocker,
\newblock J. Phys. G {\bf 38}, 035001 (2011), 1009.5239.

\bibitem{Huovinen:2012is}
P.~Huovinen and H.~Petersen,
\newblock Eur. Phys. J. A {\bf 48}, 171 (2012), 1206.3371.

\bibitem{Adamczyk:2012ku}
STAR, L.~Adamczyk {\em et~al.},
\newblock Phys. Rev. C {\bf 86}, 054908 (2012), 1206.5528.

\bibitem{Alver:2010ck}
PHOBOS, B.~Alver {\em et~al.},
\newblock Phys. Rev. C {\bf 83}, 024913 (2011), 1011.1940.

\bibitem{Arsene:2009aa}
BRAHMS, I.~Arsene {\em et~al.},
\newblock Phys. Lett. B {\bf 677}, 267 (2009), 0901.0872.

\bibitem{Appelshauser:1998yb}
NA49, H.~Appelshauser {\em et~al.},
\newblock Phys. Rev. Lett. {\bf 82}, 2471 (1999), nucl-ex/9810014.

\bibitem{Adamczyk:2017iwn}
STAR, L.~Adamczyk {\em et~al.},
\newblock Phys. Rev. C {\bf 96}, 044904 (2017), 1701.07065.

\bibitem{Abelev:2008ab}
STAR, B.~Abelev {\em et~al.},
\newblock Phys. Rev. C {\bf 79}, 034909 (2009), 0808.2041.

\bibitem{Adamczyk:2017hdl}
STAR, L.~Adamczyk {\em et~al.},
\newblock Phys. Rev. C {\bf 98}, 034918 (2018), 1701.06496.

\bibitem{Bilandzic:2010jr}
A.~Bilandzic, R.~Snellings, and S.~Voloshin,
\newblock Phys. Rev. C {\bf 83}, 044913 (2011), 1010.0233.

\bibitem{Holopainen:2010gz}
H.~Holopainen, H.~Niemi, and K.~J. Eskola,
\newblock Phys. Rev. C {\bf 83}, 034901 (2011), 1007.0368.

\bibitem{Miskowiec}
D.~Miskowiec,
\newblock \url{http://web-docs.gsi.de/~misko/overlap/}.

\bibitem{Shen:2020jwv}
C.~Shen and S.~Alzhrani,
\newblock Phys. Rev. C {\bf 102}, 014909 (2020), 2003.05852.

\bibitem{Brodsky:2012vg}
S.~Brodsky, F.~Fleuret, C.~Hadjidakis, and J.~Lansberg,
\newblock Phys. Rept. {\bf 522}, 239 (2013), 1202.6585.

\bibitem{AFTER}
J.-P. Lansberg {\em et~al.},
\newblock Adv. High Energy Phys. {\bf 2015}, 319654 (2015).

\bibitem{Hadjidakis:2018ifr}
C.~Hadjidakis {\em et~al.},
\newblock (2018), 1807.00603.

\end{thebibliography}

\end{document}